\documentclass[a4paper,12pt,reqno]{article}
\usepackage{amsmath,amsfonts,amssymb,amsthm,dsfont}
\allowdisplaybreaks

\newcommand{\Om}{\Omega}
\newcommand{\om}{\omega}
\newcommand{\half}{\tfrac{1}{2}}

\newcommand{\4}{\tilde}
\newcommand{\5}{\bar}
\newcommand{\6}{\partial}
\newcommand{\7}{\hat}

\newcommand{\Dim}{D} 

\renewcommand{\a}{a}
\renewcommand{\b}{b}
\renewcommand{\c}{c}
\newcommand{\T}{T}
\newcommand{\F}{F}
\newcommand{\E}{E}
\newcommand{\A}{A}
\newcommand{\B}{B}
\newcommand{\W}{W}
\renewcommand{\S}{S}
\newcommand{\C}{C}
\newcommand{\D}{D}
\renewcommand{\H}{H}
\newcommand{\K}{U}
\renewcommand{\SS}{S}

\newcommand{\mysection}[1]{\section{#1}
            \setcounter{equation}{0}\setcounter{figure}{0}}

\begin{document}

\begin{center}
 {\large\bfseries Consistent interactions of Curtright fields}
 \\[5mm]
 Friedemann Brandt \\[2mm]
 \textit{Institut f\"ur Theoretische Physik, Leibniz Universit\"at Hannover, Appelstra\ss e 2, 30167 Hannover, Germany}
\end{center}

\begin{abstract}
Consistent self-interactions of Curtright fields (Lorentz tensors with (2,1) Young diagram index symmetry) are constructed in  dimensions 5 and 7. Most of them modify the gauge transformations of the free theory but the commutator algebra of the deformed gauge transformations remains Abelian in all cases. All of these interactions contain terms cubic in the Curtright fields with four or five derivatives, which are reminiscent of Yang-Mills, Chapline-Manton, Freedman-Townsend and Chern-Simons interactions, respectively. 
\end{abstract}



\mysection{Introduction}\label{intro}

This work concerns consistent interactions of Curtright fields \cite{Curtright:1980yk}. Curtright fields are Lorentz tensors $T^\a_{\mu\nu\varrho}$ with Lorentz indices $\mu$,$\nu$,$\varrho$ having the permutation symmetries\footnote{Antisymmetrization of indices is defined as $X_{[\mu\nu]}=\half(X_{\mu\nu}-X_{\nu\mu})$ etc., symmetrization correspondingly as  $X_{(\mu\nu)}=\half(X_{\mu\nu}+X_{\nu\mu})$ etc.}
\begin{align}
  \T^\a_{\mu\nu\varrho}=-\T^\a_{\nu\mu\varrho}\, ,\quad  T^\a_{[\mu\nu\varrho]}=0.
	\label{eq1} 
 \end{align}
The additional index $\a$ is no Lorentz index but only enumerates the Curtright fields, i.e. we examine also models with more than one Curtright field. The Lagrangian that we use for free (non-interacting) Curtright fields is 
\begin{align}
  {\cal L}^{(0)}=-\frac 1{12}\delta_{\a\b}\,(\F^\a_{\mu\nu\varrho\sigma}\F^{\b\mu\nu\varrho\sigma}
  -3\F^\a_{\mu\nu} \F^{\b\mu\nu})
	\label{eq2}
 \end{align}
 wherein
 \begin{align}
 \F^\a_{\mu\nu\varrho\sigma}=\6_\mu \T^\a_{\nu\varrho\sigma}+\6_\nu \T^\a_{\varrho\mu\sigma}+\6_\varrho \T^\a_{\mu\nu\sigma}\, ,\quad
 \F^\a_{\mu\nu}= \F^\a_{\mu\nu\varrho}{}^\varrho
	\label{eq3}
 \end{align}
 and Lorentz indices are lowered and raised with a flat metric $\eta_{\mu\nu}$ and its inverse $\eta^{\mu\nu}$.
Curtright fields are particularly interesting in $\Dim =5$  dimensions because there a Curtright field is the elementary field
(counterpart of the metric field) in a dual formulation of linearized general relativity 
\cite{Hull:2000zn,Hull:2001iu}.

We apply the BRST-BV-cohomological approach \cite{Barnich:1993vg,Henneaux:1997bm} to construct consistent interactions. In that approach one seeks a master action $S=\SS^{(0)}+g\SS^{(1)}+g^2 \SS^{(2)}+\dots$ which solves the master equation $(S,S)=0$ \cite{Batalin:1981jr}, wherein $\SS^{(0)}$ is the master action of the original (undeformed) theory and $g$ is a deformation parameter. $S$ is thus a deformation of $\SS^{(0)}$. The master equation $(S,S)=0$ imposes $(\SS^{(0)},\SS^{(1)})=0$ at first order in $g$, $(\SS^{(1)},\SS^{(1)})+2(\SS^{(0)},\SS^{(2)})=0$ at second order etc. The first order condition $(\SS^{(0)},\SS^{(1)})=0$ requires in $\Dim$ dimensions
\begin{align}
s\om_{0,\Dim}+d\om_{1,\Dim-1}=0
	\label{eq4}
 \end{align}
wherein $s$ is the BRST differential $s\ \cdot=(\SS^{(0)},\ \cdot\ )$ of the original theory, $d=dx^\mu\6_\mu$ is the exterior derivative, $\om_{0,\Dim}$ is the integrand (exterior $\Dim$-form with ghost number 0) of $\SS^{(1)}=\int\om_{0,\Dim}$ , and $\om_{1,\Dim-1}$ is an exterior $(\Dim-1)$-form with ghost number 1 (generally $\om_{g,p}$ denotes an exterior $p$-form with ghost number $g$).

\eqref{eq4} implies descent equations $s\om_{1,\Dim-1}+d\om_{2,\Dim-2}=0$, $s\om_{2,\Dim-2}+d\om_{3,\Dim-3}=0$ etc. with increasing ghost number and decreasing form-degree that can be compactly written as (see section 9 of \cite{Barnich:2000zw} and section 3 of \cite{Dragon:2012au} for reviews)
\begin{align}
(s+d)\,\Om_\Dim=0,\quad \Om_\Dim=\sum_{p=\underline{m}}^\Dim\om_{\Dim-p,p}
	\label{eq5}
 \end{align}
 wherein $\Om_\Dim$ is a ``total form'' with ``total degree''\footnote{The total degree $G$ of a total form $\Om_G=\sum_p\om_{G-p,p}$ is the sum of the form-degree and the ghost number of its exterior forms $\om_{G-p,p}$. A total form with total degree $G$ is called a total $G$-form.} $\Dim$, and $\underline{m}$ is some form-degree at which the descent equations terminate (the value of $\underline{m}$ varies from case to case).
 
 \mysection{BRST differential}\label{brst}
 
 In our case the master action corresponding to the Lagrangian \eqref{eq2} can be taken as
 \begin{align}
\SS^{(0)}=\int [ {\cal L}^{(0)}-2(\6_\mu\S^\a_{\nu\varrho}+\6_\mu\A^\a_{\nu\varrho}
-\6_\varrho\A^\a_{\mu\nu})\T_\a^{\star\mu\nu\varrho}
-(6\S_\a^{\star\mu\nu}+2\A_\a^{\star\mu\nu})\6_\mu\C^\a_\nu ] d^\Dim x
	\label{eq6}
 \end{align}
 wherein $\S^\a_{\mu\nu}$ and $\A^\a_{\mu\nu}$ denote ghost fields, $\C^\a_\mu$ denote ghost-for-ghost fields, and $\T_\a^{\star\mu\nu\varrho}$, $\S_\a^{\star\mu\nu}$, $\A_\a^{\star\mu\nu}$ denote the antifields for $\T^\a_{\mu\nu\varrho}$, $\S^\a_{\mu\nu}$ and $\A^\a_{\mu\nu}$ respectively (the antifields for $\C^\a_\mu$ are denoted $\C_\a^{\star\mu}$). The ghost fields and antifields have the index symmetries
 \begin{align*}
& \S^\a_{\mu\nu}=\S^\a_{\nu\mu},\quad
\A^\a_{\mu\nu}=-\A^\a_{\nu\mu},\quad
\T_\a^{\star\mu\nu\varrho}=-\T_\a^{\star\nu\mu\varrho},\\
&\T_\a^{\star[\mu\nu\varrho]}=0,\quad
 \S_\a^{\star\mu\nu}=\S_\a^{\star\nu\mu},\quad
\A_\a^{\star\mu\nu}=-\A_\a^{\star\nu\mu}.
 \end{align*}
The fields, antifields, spacetime coordinates $x^\mu$ and differentials $dx^\mu$ have the following ghost numbers (gh), antifield numbers (af), Gra\ss mann parities ($|\ |$) and BRST transformations ($s$):
\begin{equation}
\begin{array}{|c|c|c|c|c|}
 \hline\rule{0em}{2.5ex}
 Z & \mathrm{gh}(Z) & \mathrm{af}(Z) & |Z| &sZ \\
 \hline\rule{0em}{2.5ex}
 \T^\a_{\mu\nu\varrho} & 0 & 0 & 0 &2 (\6_{[\mu}\S^\a_{\nu]\varrho}+\6_{[\mu}\A^\a_{\nu]\varrho}-\6_\varrho\A^\a_{\mu\nu})\\
 \hline\rule{0em}{2.5ex}
 \S^\a_{\mu\nu}& 1 & 0 & 1 & 6\6_{(\mu}\C^\a_{\nu)} \\
 \hline\rule{0em}{2.5ex}
\A^\a_{\mu\nu}& 1 & 0 & 1 & 2\6_{[\mu}\C^\a_{\nu]} \\
 \hline\rule{0em}{2.5ex}
\C^\a_{\mu}& 2 & 0 & 0 & 0 \\
 \hline\rule{0em}{2.5ex}
 \T_\a^{\star\mu\nu\varrho} & -1 & 1 & 1 & 
 \half\delta_{\a\b}\6_\sigma(\F^{\b\sigma\mu\nu\varrho}-3\F^{\b[\sigma\mu}\eta^{\nu]\varrho})=
 \delta_{\a\b}(\E^{\b\mu\nu\varrho}-\E^{\b[\mu}\eta^{\nu]\varrho})\\
 \hline\rule{0em}{2.5ex}
 \S_\a^{\star\mu\nu} & -2 & 2 & 0 & -2\6_\varrho\T_\a^{\star\varrho(\mu\nu)}\\
  \hline\rule{0em}{2.5ex}
  \A_\a^{\star\mu\nu} & -2 & 2 & 0 & 3\6_\varrho\T_\a^{\star\mu\nu\varrho}=-6\6_\varrho\T_\a^{\star\varrho[\mu\nu]}\\
  \hline\rule{0em}{2.5ex}
  \C_\a^{\star\mu} & -3 & 3 & 1 & \6_\nu (6\S_\a^{\star\nu\mu}+2\A_\a^{\star\nu\mu})\\
  \hline\rule{0em}{2.5ex}
  x^\mu & 0 & 0 & 0 & 0\\
    \hline\rule{0em}{2.5ex}
  dx^\mu & 0 & 0 & 1 & 0\\
 \hline
 \end{array}
 \label{eq7}
\end{equation}
wherein $\E^\a_{\mu\nu\varrho}$ and $\E^\a_\mu$ are traces of a gauge invariant tensor $\E^\a_{\mu\nu\varrho\sigma\tau}$:
 \begin{align}
\E^\a_{\mu\nu\varrho\sigma\tau}=\half(\6_\sigma\F^\a_{\mu\nu\varrho\tau}-\6_\tau\F^\a_{\mu\nu\varrho\sigma}),
\ \E^\a_{\mu\nu\varrho}=-\E^{\a}_{\sigma\mu\nu\varrho}{}^\sigma,\
\E^\a_{\mu}=\E^\a_{\mu\nu}{}^\nu.
	\label{eq8}
 \end{align}
 These tensors fulfill the identities
   \begin{align}
&\E^\a_{[\mu\nu\varrho\sigma]\tau}=0,\ 
\E^\a_{[\mu\nu\varrho]}=0,
	\label{eq9}\\
&\6^\tau\E^\a_{\mu\nu\varrho\sigma\tau}=-3\6_{[\mu}\E^\a_{\nu\varrho]\sigma}\,,\
\6^\tau\E^\a_{\tau\mu\nu\varrho\sigma}=\6_{\varrho}\E^\a_{\mu\nu\sigma}-\6_{\sigma}\E^\a_{\mu\nu\varrho}\,,
	\label{eq10}\\
&\6^\mu\E^\a_{\mu\nu\varrho}=-\half\6_{\varrho}\E^\a_{\nu}\,,\
\6^\varrho\E^\a_{\mu\nu\varrho}=-\6_{[\mu}\E^\a_{\nu]}\, .\label{eq11}
 \end{align}
 For later purpose we also introduce the totally tracefree part $\W^\a_{\mu\nu\varrho\sigma\tau}$ of $\E^\a_{\mu\nu\varrho\sigma\tau}$ in dimensions $\Dim>3$:
  \begin{align}
  \W^\a_{\mu\nu\varrho}{}^{\sigma\tau}=
\E^\a_{\mu\nu\varrho}{}^{\sigma\tau}
+\tfrac 6{\Dim-3}\E^\a_{[\mu\nu}{}^{[\sigma}\delta^{\tau]}_{\varrho]}
-\tfrac 6{(\Dim-3)(\Dim-2)}\E^a_{[\mu}\delta^\sigma_{\nu}\delta^\tau_{\varrho]}\,.
	\label{eq12}
 \end{align}
 We remark that $\F^\a_{\mu\nu\varrho\sigma}$, $\E^\a_{\mu\nu\varrho\sigma\tau}$, $\W^\a_{\mu\nu\varrho\sigma\tau}$, $\E^\a_{\mu\nu\varrho}$ and $\E^\a_\mu$ are the counterparts of the linearized Levi-Civita-Christoffel connection, Riemann-Christoffel tensor, Weyl tensor, Ricci tensor and curvature scalar of general relativity, respectively. $\E^\a_{\mu\nu\varrho}$ and $\E^\a_\mu$ vanish on-shell in the free theory, and $\E^\a_{\mu\nu\varrho\sigma\tau}$ equals $\W^\a_{\mu\nu\varrho\sigma\tau}$ on-shell in the free theory:
   \begin{align}
   &\E^{\a\mu}=-\tfrac{2}{\Dim-3}s\T_\b^{\star\mu}\delta^{\b\a}\approx 0,\label{eq12a}\\
   &\E^{\a\mu\nu\varrho}
  =s(\T_\b^{\star\mu\nu\varrho}-\tfrac{2}{\Dim-3}\T_\b^{\star[\mu}\eta^{\nu]\varrho})\delta^{\b\a}
  \approx 0,\label{eq12b}\\
  &\E^{\a\mu\nu\varrho}{}_{\sigma\tau}= \W^{\a\mu\nu\varrho}{}_{\sigma\tau}
  -\tfrac{6}{\Dim-3}s(\T_\b^{\star[\mu\nu}{}_{[\sigma}\delta^{\varrho]}_{\tau]}
  -\tfrac{2}{\Dim-2}\T_\b^{\star[\mu}\delta^{\nu}_\sigma\delta^{\varrho]}_\tau)\delta^{\b\a}
  \approx \W^{\a\mu\nu\varrho}{}_{\sigma\tau} 
	\label{eq12c}
 \end{align}
 wherein 
 \begin{align}
  \T_{\a}^{\star\mu}=\T_\a^{\star\mu\nu}{}_\nu
	\label{eq12d}
 \end{align}
 and $\approx$ denotes equality on-shell in the free theory ($s \T_\a^{\star\mu\nu\varrho}$ is the Euler-Lagrange derivative of ${\cal L}^{(0)}$ with respect to $\T^\a_{\mu\nu\varrho}$, i.e. the BRST-transformations 
 $s \T_\a^{\star\mu\nu\varrho}$ are the 
 ``left hand sides'' of the equations of motion of the free theory). 
 
 \mysection{Constituent total forms}\label{cons}
 
 To construct solutions of equations \eqref{eq4} and \eqref{eq5} we define total 1-forms $\Om_1^{\a\mu\nu\varrho}$ and 2-forms $\Om_2^{\a\mu\nu\varrho}$:
 \begin{align}
 \Om_1^{\a\mu\nu\varrho}=\H^{\a\mu\nu\varrho}-\F^{\a\mu\nu\varrho}{}_\sigma dx^\sigma,\quad
 \Om_2^{\a\mu\nu\varrho}=-\E^{\a\mu\nu\varrho}{}_{\sigma\tau} dx^\sigma dx^\tau\label{eq13a}
 \end{align}
 wherein
 \begin{align}
 \H^\a_{\mu\nu\varrho}=6\6_{[\mu}\A^\a_{\nu\varrho]}\,.\label{eq13b}
 \end{align}
 The forms defined in equations \eqref{eq13a} fulfill
  \begin{align}
(s+d)\,\Om_1^{\a\mu\nu\varrho}=\Om_2^{\a\mu\nu\varrho},\quad (s+d)\,\Om_2^{\a\mu\nu\varrho}=0.\label{eq13c}
 \end{align}
Furthermore in dimensions $\Dim>4$ we define total $(\Dim-3)$-forms $\Om_{\Dim-3}^{\a\mu}$:
 \begin{align}
 &\Om_{\Dim-3}^{\a\mu}=\sum_{p=\Dim-3}^\Dim\om_{\Dim-3-p,p}^{\a\mu}\,,\nonumber\\
  &\om_{-3,\Dim}^{\a\mu}=\delta^{ab}\C_\b^{\star\mu} d^\Dim x,\nonumber\\
  &\om_{-2,\Dim-1}^{\a\mu}=-\delta^{ab}(6\S_\b^{\star\nu\mu}+2\A_\b^{\star\nu\mu})(d^{\Dim-1}x)_\nu\,,\nonumber\\
  &\om_{-1,\Dim-2}^{\a\mu}=\tfrac{36}{\Dim-3}\delta^{ab}
  (x^{[\nu}\6_{\sigma}\T_\b^{\star\varrho\sigma]\mu}
  -x^\tau\eta^{\mu[\nu}\6_{\sigma}\T_\b^{\star\varrho\sigma]}{}_\tau
  -\tfrac{2}{\Dim-2}\eta^{\mu[\nu}x^{\varrho}\6_{\sigma}\T_\b^{\star\sigma]})
  (d^{\Dim-2}x)_{\nu\varrho}\,,\nonumber\\
   &\om_{0,\Dim-3}^{\a\mu}=-\tfrac{12}{\Dim-4}\, \W^{\a\nu\varrho\sigma\mu}{}_\tau x^\tau 
   (d^{\Dim-3}x)_{\nu\varrho\sigma}\label{eq13d}
 \end{align}
 and total $(\Dim-2)$-forms $\Om_{\Dim-2}^{\a\mu\nu\varrho}$:
 \begin{align}
 &\Om_{\Dim-2}^{\a\mu\nu\varrho}=\om_{-1,\Dim-1}^{\a\mu\nu\varrho}+\om_{0,\Dim-2}^{\a\mu\nu\varrho}\, ,\nonumber\\
 &\om_{-1,\Dim-1}^{\a\mu\nu\varrho}=3\delta^{ab}(\6^{[\mu}\T_\b^{\star\nu\varrho]\sigma}
 -\eta^{\sigma[\mu}\6_\tau\T_\b^{\star\nu\varrho]\tau}
 -\tfrac{2}{\Dim-3}\eta^{\sigma[\mu}\6^{\nu}\T_\b^{\star\varrho]})
 (d^{\Dim-1}x)_{\sigma}\, ,\nonumber\\
 &\om_{0,\Dim-2}^{\a\mu\nu\varrho}=\E^{\a\mu\nu\varrho\sigma\tau}(d^{\Dim-2}x)_{\sigma\tau}\label{eq13e}
 \end{align}
 wherein 
  \begin{align}
&(d^{\Dim-p}x)_{\mu_1\ldots\mu_p}=\tfrac 1{(\Dim-p)!p!}\,\epsilon_{\mu_1\ldots\mu_\Dim}dx^{\mu_{p+1}}\ldots dx^{\mu_{\Dim}}\,.\label{eq14}
 \end{align}
 The forms defined in equations \eqref{eq13d} and \eqref{eq13e} fulfill
  \begin{align}
  (s+d)\,\Om_{\Dim-3}^{\a\mu}=0,\quad
(s+d)\,\Om_{\Dim-2}^{\a\mu\nu\varrho}=0.\label{eq13f}
 \end{align}
 
{\bf Comments:}
 
 (i) The total $(\Dim-3)$-forms $\Om_{\Dim-3}^{\a\mu}$ defined in equations \eqref{eq13d} derive from the following
 simpler total $(\Dim-3)$-forms $\Lambda_{\Dim-3}^{\a\mu}$:
 \begin{align}
 &\Lambda_{\Dim-3}^{\a\mu}=\om_{-3,\Dim}^{\a\mu}+\om_{-2,\Dim-1}^{\a\mu}+\lambda_{-1,\Dim-2}^{\a\mu}
 +\lambda_{0,\Dim-3}^{\a\mu}+\lambda_{1,\Dim-4}^{\a\mu}\,,\nonumber\\
  &\lambda_{-1,\Dim-2}^{\a\mu}=-12\delta^{ab}\T_\b^{\star\nu_1\nu_2\mu}(d^{\Dim-2}x)_{\nu_1\nu_2}\,,\nonumber\\
   &\lambda_{0,\Dim-3}^{\a\mu}=6(\F^{\a\nu_1\nu_2\nu_3\mu}
   -3\eta^{\mu\nu_1}\F^{\a\nu_2\nu_3\varrho}{}_{\varrho})(d^{\Dim-3}x)_{\nu_1\nu_2\nu_3}\,,\nonumber\\
      &\lambda_{1,\Dim-4}^{\a\mu}=24 \eta^{\mu\nu_1}\H^{\a\nu_2\nu_3\nu_4}(d^{\Dim-4}x)_{\nu_1\ldots\nu_4}\label{eq13}
 \end{align}
 with $\om_{-3,\Dim}^{\a\mu}$ and $\om_{-2,\Dim-1}^{\a\mu}$ as in equations \eqref{eq13d}.
 
 Using table \eqref{eq7} it can be readily checked that the total forms $\Lambda_{\Dim-3}^{\a\mu}$ are $(s+d)$-cocycles:
 \begin{align}
 (s+d)\Lambda_{\Dim-3}^{\a\mu}=0.
\label{eq17}
 \end{align}
 Furthermore it can readily be shown that $\Lambda_{\Dim-3}^{\a\mu}$ is no $(s+d)$-coboundary. Indeed, $\Lambda_{\Dim-3}^{\a\mu}=(s+d)\eta_{\Dim-4}^{\a\mu}$ would imply $\om_{-3,\Dim}^{\a\mu}=\delta^{ab}\C_\b^{\star\mu} d^\Dim x=s\eta^{\a\mu}_{-4,\Dim}+d\eta^{\a\mu}_{-3,\Dim-1}$ for some 
 local exterior forms $\eta^{\a\mu}_{-4,\Dim}$ and $\eta^{\a\mu}_{-3,\Dim-1}$ which can be easily shown not to exist. 
 Hence, $\Lambda_{\Dim-3}^{\a\mu}$ is nontrivial in the cohomology of $(s+d)$.
 
 $\lambda_{0,\Dim-3}^{\a\mu}$ in $\Lambda_{\Dim-3}^{\a\mu}$ is a conserved exterior $(\Dim-3)$-form of the free theory because \eqref{eq17} contains
 \begin{align*}
 d\lambda_{0,\Dim-3}^{\a\mu}=-s\lambda_{-1,\Dim-2}^{\a\mu}\approx 0.
 \end{align*}
 Now, $\lambda_{0,\Dim-3}^{\a\mu}$ is not gauge invariant in the free theory because \eqref{eq17} also contains the equation
 \begin{align*}
 s\lambda_{0,\Dim-3}^{\a\mu}=-d\lambda_{1,\Dim-4}^{\a\mu}\neq 0.
 \end{align*}
 However, for $\Dim>4$ we can ``improve'' $\Lambda_{\Dim-3}^{\a\mu}$ by subtracting an $(s+d)$-coboundary from it which removes the exterior $(\Dim-4)$-form from it and makes the resulting exterior $(\Dim-3)$-form gauge invariant. Indeed, we have
  \begin{align}
 \Dim>4:\ \lambda_{1,\Dim-4}^{\a\mu}=d\eta^{\a\mu}_{1,\Dim-5}+s\eta^{\a\mu}_{0,\Dim-4}
 \label{eq18}
 \end{align}
 wherein
  \begin{align}
\eta^{\a\mu}_{1,\Dim-5}&=\tfrac{120}{\Dim-4}\,x^{\nu_5}\H^{\a\nu_4\nu_3\nu_2}\eta^{\nu_1\mu}
(d^{\Dim-5}x)_{\nu_1\ldots\nu_5}\, ,
 \label{eq19}\\
 \eta^{\a\mu}_{0,\Dim-4}&=\tfrac{24}{\Dim-4}\,(\F^{\a\nu_4\nu_3\nu_2\mu}x^{\nu_1}-
 4\F^{\a[\nu_4\nu_3\nu_2}{}_\varrho x^{\varrho]}\eta^{\nu_1\mu})(d^{\Dim-4}x)_{\nu_1\ldots\nu_4}\, .
  \label{eq20}
 \end{align}
\eqref{eq18} implies that the total $(\Dim-3)$-form $\Lambda_{\Dim-3}^{\a\mu}-(s+d)(\eta^{\a\mu}_{1,\Dim-5}+\eta^{\a\mu}_{0,\Dim-4})$ contains only exterior $p$-forms with form-degrees $p\geq\Dim-3$.\footnote{This 
total $(\Dim-3)$-form very likely coincides with the total $(\Dim-3)$ form $\4{\cal H}_\mu$ of \cite{Bekaert:2004dz} that occurs there in the case $(p,q)=(2,1)$, see section 3.3 of the arXiv-version of \cite{Bekaert:2004dz}.}
Moreover its exterior $(\Dim-3)$-form is
 \begin{align}
& \lambda^{\a\mu}_{0,\Dim-3}-d\eta^{\a\mu}_{0,\Dim-4}=
 -\tfrac{12}{\Dim-4}\,K^{\a\nu_3\nu_2\nu_1\mu}
(d^{\Dim-3}x)_{\nu_1\nu_2\nu_3}\,,\nonumber\\
&K^\a_{\nu_3\nu_2\nu_1\mu}=\E^\a_{\nu_3\nu_2\nu_1\mu\varrho} x^\varrho
  +3x_{[\nu_3}\E^{\a}_{\nu_2\nu_1]\mu}
  +3\eta_{\mu[\nu_3}\E^\a_{\nu_2}x_{\nu_1]}
  -3\eta_{\mu[\nu_3}\E^\a_{\nu_2\nu_1]\varrho} x^\varrho.
 \label{eq21}
 \end{align}
 Notice that this exterior $(\Dim-3)$-form indeed is gauge invariant and that it does not contain any $x$-independent terms. 
 In fact, the $x$-independent terms of $d\eta^{\a\mu}_{0,\Dim-4}$ cancel exactly $\lambda^{\a\mu}_{0,\Dim-3}$ and only the $x$-dependent terms of $d\eta^{\a\mu}_{0,\Dim-4}$ survive. \eqref{eq21} is already a gauge invariant improvement of $\lambda^{\a\mu}_{0,\Dim-3}$ but we proceed one step further and remove also terms from the 
 exterior $(\Dim-3)$-form \eqref{eq21} which vanish on-shell in the free theory. Using equations \eqref{eq12a}-\eqref{eq12c}
 one finds that such terms are the BRST-transformation of the following exterior $(\Dim-3)$-form $\eta^{\a\mu}_{-1,\Dim-3}$:
   \begin{align}
\eta^{\a\mu}_{-1,\Dim-3}&=\tfrac{36}{\Dim-3}\,\delta^{\a\b}(\eta^{\mu\nu_3}\T_\b^{\star\nu_2\nu_1}{}_\varrho x^\varrho
-x^{\nu_3}\T_\b^{\star\nu_2\nu_1\mu}+\tfrac{2}{\Dim-2}\,\eta^{\mu\nu_3}x^{\nu_2}\T_\b^{\star\nu_1})(d^{\Dim-3}x)_{\nu_1\nu_2\nu_3}\, .
  \label{eq22}
 \end{align}
  We arrive at the improved total form \eqref{eq13d}:
    \begin{align}
\Lambda_{\Dim-3}^{\a\mu}-(s+d)(\eta^{\a\mu}_{1,\Dim-5}+\eta^{\a\mu}_{0,\Dim-4}+\eta^{\a\mu}_{-1,\Dim-3})
=\Om_{\Dim-3}^{\a\mu}\,.
  \label{eq23}
  \end{align}
  $(s+d)\Om_{\Dim-3}^{\a\mu}=0$ is thus a direct consequence of \eqref{eq17}.  As $\Lambda_{\Dim-3}^{\a\mu}$ is nontrivial in the cohomology of $(s+d)$, $\Om_{\Dim-3}^{\a\mu}$ is also nontrivial in that cohomology.
  
  Notice also that the exterior $(\Dim-2)$-form $\om^{\a\mu}_{-1,\Dim-2}$ in $\Om_{\Dim-3}^{\a\mu}$ does not contain any
  $x$-independent terms either. This parallels what happened for the exterior $(\Dim-3)$-form: 
  the $x$-independent terms of $d\eta^{\a\mu}_{-1,\Dim-3}$ cancel exactly $\lambda^{\a\mu}_{-1,\Dim-2}$ 
  and only the $x$-dependent terms of $d\eta^{\a\mu}_{-1,\Dim-3}$ survive in $\om^{\a\mu}_{-1,\Dim-2}$. 
  In fact one can proceed further and remove also the $s$-trivial terms in $\om^{\a\mu}_{-1,\Dim-2}$ 
  (i.e. the terms with $\6_{\sigma}\T_\b^{\star\sigma\cdot\cdot}$ and $\6_{\sigma}\T_\b^{\star\sigma}$) by subtracting
  a total form $(s+d)\eta^{\a\mu}_{-2,\Dim-2}$ from $\Lambda_{\Dim-3}^{\a\mu}$, and afterwards also the $s$-trivial terms in the resultant redefined exterior $(\Dim-1)$-form and exterior $\Dim$-form which however appears to be merely of academic interest and therefore is not done here (the exterior $p$-forms with $p>\Dim-2$ in $\Om_{\Dim-3}^{\a\mu}$ anyway do not contribute to the deformations constructed below). 
  
 We remark that it is impossible to improve $\Lambda_{\Dim-3}^{\a\mu}$ to an $(s+d)$-cocycle with a gauge invariant and $x$-independent exterior $(\Dim-3)$-form. Indeed, such an improvement would require the existence of $x$-independent exterior forms $\eta^{\a\mu}_{0,\Dim-4}$ and $\eta^{\a\mu}_{1,\Dim-5}$ that fulfill \eqref{eq18} but it can 
 easily be shown that such forms do not exist. The improvement of $\Lambda_{\Dim-3}^{\a\mu}$ thus necessarily depends explicitly on the coordinates $x$. Furthermore the improvement is crucial for the construction of consistent deformations involving $\Om_{\Dim-3}^{\a\mu}$, as will become clear below. 
 
 (ii) The total $(\Dim-2)$-forms $\Om_{\Dim-2}^{\a\mu\nu\varrho}$ defined in equations \eqref{eq13e} are actually 
 $(s+d)$-exact, i.e. one has $\Om_{\Dim-2}^{\a\mu\nu\varrho}=(s+d)\eta_{\Dim-3}^{\a\mu\nu\varrho}$ 
 for some total $(\Dim-3)$-form $\eta_{\Dim-3}^{\a\mu\nu\varrho}$. 
 This follows already from the fact that $\Om_{\Dim-2}^{\a\mu\nu\varrho}$ has no exterior $\Dim$-form. In particular
 the exterior $(\Dim-2)$-form $\om_{0,\Dim-2}^{\a\mu\nu\varrho}$ of $\Om_{\Dim-2}^{\a\mu\nu\varrho}$ is thus
 trivial, i.e. $\om_{0,\Dim-2}^{\a\mu\nu\varrho}=d\eta_{0,\Dim-3}^{\a\mu\nu\varrho}+s\eta_{-1,\Dim-2}^{\a\mu\nu\varrho}$ for some exterior $(\Dim-3)$-form $\eta_{0,\Dim-3}^{\a\mu\nu\varrho}$ and 
 some exterior $(\Dim-2)$-form $\eta_{-1,\Dim-2}^{\a\mu\nu\varrho}$. In other words, $\om_{0,\Dim-2}^{\a\mu\nu\varrho}$ is $d$-exact on-shell in the free theory. However, it is not $d$-exact on-shell in the space of gauge invariant
 and $x$-independent exterior forms, i.e. there is no 
 gauge invariant exterior $(\Dim-3)$-form $\eta_{0,\Dim-3}^{\a\mu\nu\varrho}$ 
 which does not dependent explicitly on the coordinates $x$ such that 
 $\om_{0,\Dim-2}^{\a\mu\nu\varrho}\approx d\eta_{0,\Dim-3}^{\a\mu\nu\varrho}$ (e.g. in $\Dim=5$ 
 one has $\om_{0,3}^{\a\mu\nu\varrho}\approx d\eta_{0,2}^{\a\mu\nu\varrho}$ with
 $\eta_{0,2}^{\a\mu_1\mu_2\mu_3}\propto \epsilon^{\mu_1\ldots\mu_5}
 \F^\a_{\mu_4\mu_5\nu\varrho}dx^\nu dx^\varrho$).
 
 (iii) $\Om_2^{\a\mu\nu\varrho}$ in \eqref{eq13a} and $\Om_{\Dim-2}^{\a\mu\nu\varrho}$ in \eqref{eq13e} can also be modified by removing terms that vanish on-shell in the free theory. In particular, 
using equation \eqref{eq12c} one can write terms of $\Om_2^{\a\mu\nu\varrho}$  that vanish on-shell in the free theory as 
$s\eta_{-1,2}^{\a\mu\nu\varrho}$ with an exterior 2-form $\eta_{-1,2}^{\a\mu\nu\varrho}$ and redefine $\Om_{1}^{\a\mu\nu\varrho}\rightarrow
\Om_{1}^{\a\mu\nu\varrho}-\eta_{-1,2}^{\a\mu\nu\varrho}$ and  $\Om_{2}^{\a\mu\nu\varrho}\rightarrow
\Om_{2}^{\a\mu\nu\varrho}-(s+d)\eta_{-1,2}^{\a\mu\nu\varrho}=-\W^{\a\mu\nu\varrho}{}_{\sigma\tau} dx^\sigma dx^\tau-d\eta_{-1,2}^{\a\mu\nu\varrho}$. 
Correspondingly one can write terms of $\om_{0,\Dim-2}^{\a\mu\nu\varrho}$  that vanish on-shell in the free theory as 
$s\eta_{-1,\Dim-2}^{\a\mu\nu\varrho}$ with an exterior $(\Dim-2)$-form $\eta_{-1,\Dim-2}^{\a\mu\nu\varrho}$ (proportional to the Hodge dual of $\eta_{-1,2}^{\a\mu\nu\varrho}$) and redefine $\Om_{\Dim-2}^{\a\mu\nu\varrho}\rightarrow
\Om_{\Dim-2}^{\a\mu\nu\varrho}-(s+d)\eta_{-1,\Dim-2}^{\a\mu\nu\varrho}$ whose exterior $(\Dim-2)$-form
is $\W^{\a\mu\nu\varrho\sigma\tau}(d^{\Dim-2}x)_{\sigma\tau}$. The first order consistent deformations constructed below from 
 $\Om_{1}^{\a\mu\nu\varrho}$, $\Om_{2}^{\a\mu\nu\varrho}$ and $\Om_{\Dim-2}^{\a\mu\nu\varrho}$
 can be constructed likewise (and equivalently) with the redefined total forms.

\mysection{Consistent first order deformations}\label{first}
  
 Using the total forms \eqref{eq13a}, \eqref{eq13d} and \eqref{eq13e} we now construct solutions of equation
 \eqref{eq5} in dimensions $\Dim=5$ and $\Dim=7$ which are cubic in the fields and antifields and which we call ``Yang-Mills type'', ``Chapline-Manton type'', ``Freedman-Townsend type'' and ``Chern-Simons type'' solutions (this wording will be justified in section \ref{conclusion}), and which we denote $\Om^{\mathrm{YM}}_\Dim$, 
  $\Om^{\mathrm{CM}}_\Dim$,  $\Om^{\mathrm{FT}}_\Dim$ and  $\Om^{\mathrm{CS}}_\Dim$, respectively.

The Yang-Mills type solutions are:
\begin{align}
\Dim=5:\quad & \Om^{\mathrm{YM}}_5=\epsilon_{\mu_1\ldots\mu_5}\Om_3^{\a\mu_1\mu_2\mu_3}
\Om_1^{\b\mu_4\nu\varrho}\Om_1^{\c\mu_5}{}_{\nu\varrho}f_{\a\b\c}\,, \label{ym5}\\
\Dim=7:\quad &\Om^{\mathrm{YM}}_7=\epsilon_{\mu_1\ldots\mu_7}\Om_5^{\a\mu_1\mu_2\mu_3}
\Om_1^{\b\mu_4\mu_5\nu}\Om_1^{\c\mu_6\mu_7}{}_{\nu}f_{\a\b\c}\label{ym7}
  \end{align}
  wherein $f_{\a\b\c}$ are constant coefficients that are totally symmetric in $\Dim=5$ and 
  totally antisymmetric in $\Dim=7$ (and otherwise arbitrary, at least at first order):
  \begin{align}
\Dim=5:\ 
f_{\a\b\c}=f_{(\a\b\c)}\,, \quad
\Dim=7:\
f_{\a\b\c}=f_{[\a\b\c]}\,. \label{f}
  \end{align}
  In \eqref{ym5} and \eqref{ym7} $\Om_{3}^{\a\cdots}$ and $\Om_5^{\a\cdots}$ are the total $(\Dim-2)$-forms of \eqref{eq13e} for $\Dim=5$ and $\Dim=7$, and 
  $\Om_1^{\b\cdots}$ and $\Om_1^{\c\cdots}$ are the total 1-forms of \eqref{eq13a}.

The Chapline-Manton type solutions are:
\begin{align}
\Dim=5:\quad & \Om^{\mathrm{CM}}_5=\epsilon_{\mu_1\ldots\mu_5}\Om_{2}^{\a\nu}
\Om_2^{\b\mu_1\mu_2\mu_3}\Om_1^{\c\mu_4\mu_5}{}_\nu\, e_{\a\b\c}\,,\label{cm5}\\
\Dim=7:\quad &\Om^{\mathrm{CM}}_7=\epsilon_{\mu_1\ldots\mu_7}\Om_4^{\a\mu_1}
\Om_2^{\b\mu_2\mu_3\mu_4}
\Om_1^{\c\mu_5\mu_6\mu_7}e_{\a\b\c}\label{cm7}
  \end{align}
wherein $e_{\a\b\c}$ are constant coefficients that are symmetric in $\Dim=5$  and 
antisymmetric symmetric in $\Dim=7$ in the last two indices (and otherwise arbitrary):
  \begin{align}
\Dim=5:\ 
e_{\a\b\c}=e_{\a\c\b}\,, \quad
\Dim=7:\
e_{\a\b\c}=-e_{\a\c\b}\,. \label{e}
  \end{align}
   In \eqref{cm5} and \eqref{cm7} $\Om_{2}^{\a\cdot}$ and $\Om_{4}^{\a\cdot}$ 
   are the total $(\Dim-3)$-forms of \eqref{eq13d} for $\Dim=5$ and $\Dim=7$, and 
  $\Om_1^{\c\cdots}$ and $\Om_2^{\b\cdots}$ are the total 1-forms and 2-forms of \eqref{eq13a}.
  
Cubic  Freedman-Townsend and Chern-Simons type solutions exist only in $\Dim=5$ dimensions:
\begin{align}
\Dim=5:\quad &\Om^{\mathrm{FT}}_5=\epsilon_{\mu_1\ldots\mu_5}
\Om_{2}^{\a\mu_1}\Om_{2}^{\b\mu_2}
\Om_1^{\c\mu_3\mu_4\mu_5}d_{\a\b\c}\,,\label{ft5}\\
& \Om^{\mathrm{CS}}_{5}=\epsilon_{\mu_1\ldots\mu_5}\Om_1^{\a\mu_1\mu_2\mu_3}
\Om_2^{\b\mu_4\nu\varrho}\Om_2^{\c\mu_5}{}_{\nu\varrho}\,c_{\a\b\c}\label{cs5}
  \end{align}
 wherein $d_{\a\b\c}$ are constant coefficients that are antisymmetric in the first two indices and $c_{\a\b\c}$ are totally antisymmetric constant coefficients (otherwise these coefficients are arbitrary): 
  \begin{align}
d_{\a\b\c}=-d_{\b\a\c}\,,\quad
c_{\a\b\c}=c_{[\a\b\c]}\,. \label{cd}
  \end{align}
  In \eqref{ft5} $\Om_{2}^{\a\cdot}$ and $\Om_{2}^{\b\cdot}$ 
   are the total $(\Dim-3)$-forms of \eqref{eq13d} for $\Dim=5$ and $\Om_1^{\c\cdots}$ are the total 1-forms of \eqref{eq13a}, and in \eqref{cs5} $\Om_1^{\a\cdots}$ are the total 1-forms of \eqref{eq13a}, and $\Om_2^{\b\cdots}$ and $\Om_2^{\c\cdots}$ are the 2-forms of \eqref{eq13a}. 
  
 {\bf Comments:}
 
 (i) The total $5$-forms $\Om^{\mathrm{CM}}_5$, $\Om^{\mathrm{FT}}_5$ and $\Om^{\mathrm{CS}}_5$ given in equations \eqref{cm5}, \eqref{ft5} and \eqref{cs5} solve equation \eqref{eq5} 
 because $(s+d)\Om^{\mathrm{CM}}_5$, $(s+d)\Om^{\mathrm{FT}}_5$ 
 and $(s+d)\Om^{\mathrm{CS}}_5$ contain only exterior forms with form-degrees $p>5$ and thus vanish in $\Dim=5$, as can be readily checked.\footnote{For this result it is crucial that the exterior $(\Dim-3)$-form
 $\om_{0,\Dim-3}^{\a\mu}$ of
 the total $(\Dim-3)$-form $\Om_{\Dim-3}^{\a\mu}$ given in \eqref{eq13d} is gauge invariant because otherwise 
$\Om_{\Dim-3}^{\a\mu}$ would contain an exterior $(\Dim-4)$-form and the reasoning for $\Om^{\mathrm{CM}}_5$ and $\Om^{\mathrm{FT}}_5$ would fail. This likewise applies to $\Om^{\mathrm{CM}}_7$.}
 Similarly the total 7-form $\Om^{\mathrm{CM}}_7$ given in equation \eqref{cm7} solves equation \eqref{eq5} 
 because $(s+d)\Om^{\mathrm{CM}}_7$ contains only exterior forms with 
 form-degrees $p>7$ and thus vanishes in $\Dim=7$. The symmetries \eqref{e} and \eqref{cd} of the coefficients 
 $e_{\a\b\c}$ and $c_{\a\b\c}$ avoid that the total forms
 $\Om^{\mathrm{CM}}_5$, $\Om^{\mathrm{CM}}_7$ and $\Om^{\mathrm{CS}}_5$
 are obviously $(s+d)$-exact (the symmetry of the $e$'s avoids that
 $\Om^{\mathrm{CM}}_\Dim$ has the structure $(s+d)(\Om_{\Dim-3}\Om_1\Om_1)$, the 
 symmetry of the $c$'s avoids that $\Om^{\mathrm{CS}}_{5}$ has the structure $(s+d)(\Om_{2}\Om_1\Om_1)$).
 The antisymmetry \eqref{cd} of the coefficients 
 $d_{\a\b\c}$ simply reflects the even Gra\ss mann parity of the total forms $\Om_{2}^{\a\mu}$
 and the antisymmetry of $\epsilon_{\mu_1\ldots\mu_5}$.

 (ii)  $(s+d)\Om^{\mathrm{YM}}_\Dim=0$ for $\Dim=5$ and $\Dim=7$ can be shown as follows. For an object 
 $Z_{\varrho_1\varrho_2\varrho_3}=Z_{[\varrho_1\varrho_2\varrho_3]}$ in $\Dim=2k+1$ dimensions we define
 \begin{align}
\Dim=2k+1:\ \4Z^{\mu_1\ldots\mu_{k-1}\nu_1\ldots\nu_{k-1}}:=
\epsilon^{\mu_1\ldots\mu_{k-1}\nu_1\ldots\nu_{k-1}\varrho_1\varrho_2\varrho_3}
Z_{\varrho_1\varrho_2\varrho_3}\equiv \4Z^{(\mu)(\nu)}
\label{tilde1}
  \end{align}
  where $(\mu)$ and $(\nu)$ denote the multi-indices $[\mu_1\ldots\mu_{k-1}]$ and
  $[\nu_1\ldots\nu_{k-1}]$, respectively (in $\Dim=5$ $(\mu)$ and $(\nu)$ are not multi-indices but just ordinary indices).
   Notice that
   \begin{align}
  \4Z^{(\mu)(\nu)}=(-)^{k-1}\4Z^{(\nu)(\mu)}.
  \label{tilde2}
  \end{align}
  With this multi-index notation the total forms $\Om^{\mathrm{YM}}_\Dim$ in equations \eqref{ym5} and \eqref{ym7}
  can be written as
  \begin{align}
\Om^{\mathrm{YM}}_{2k+1}\propto \4\Om_{2k-1}^{\a(\mu)(\nu)} \4\Om^\b_{1(\varrho)(\mu)}
\4\Om_{1}^{\c(\varrho)}{}_{(\nu)}f_{\a\b\c}
\label{tilde3}
  \end{align}
  and one obtains, using $(s+d)\4\Om_1^{\a(\mu)(\nu)}=\4\Om_2^{\a(\mu)(\nu)}$ 
  (which holds owing to \eqref{eq13c}):
   \begin{align}
(s+d)\Om^{\mathrm{YM}}_{2k+1}&\propto \4\Om_{2k-1}^{\a(\mu)(\nu)} \4\Om^\b_{2(\varrho)(\mu)}
\4\Om_{1}^{\c(\varrho)}{}_{(\nu)}f_{\a\b\c}\nonumber\\
&=- d^{2k+1}x\, \4\E^{\a(\mu)(\nu)\sigma\tau}\4\E^\b_{(\varrho)(\mu)\sigma\tau}\4\H^{\c(\varrho)}{}_{(\nu)}f_{\a\b\c}
\label{tilde4}
  \end{align}
  where we used that $\4\Om_{2k-1}^{\a(\mu)(\nu)} \4\Om^\b_{2(\varrho)(\mu)}=- d^{2k+1}x\, \4\E^{\a(\mu)(\nu)\sigma\tau}\4\E^\b_{(\varrho)(\mu)\sigma\tau}$ is an exterior volume form
 which is implied by equations \eqref{eq13a} and \eqref{eq13e}. \eqref{tilde4} vanishes because of 
  \eqref{tilde2} if $f_{\a\b\c}=f_{(\a\b\c)}$ for $k=2m$ and $f_{\a\b\c}=f_{[\a\b\c]}$ for $k=2m+1$. We remark that
  \eqref{tilde3} actually vanishes for $k>4$ because in dimensions $\Dim=2k+1>9$ there is no way to contract
  the nine free Lorentz indices of $\Om_{2k-1}^{\a\mu_1\mu_2\mu_3}
\Om_1^{\b\mu_4\mu_5\mu_6}\Om_1^{\c\mu_7\mu_8\mu_9}$ in a Lorentz invariant way. For the same reason there is
no $\Om^{\mathrm{YM}}_{\Dim}$ in even dimensions $\Dim$. In $\Dim=9$ one obtains
\begin{align*}
\Dim=9:\ \Om^{\mathrm{YM}}_{9}=\epsilon_{\mu_1\ldots\mu_9}\Om_7^{\a\mu_1\mu_2\mu_3}
\Om_1^{\b\mu_4\mu_5\mu_6}\Om_1^{\c\mu_7\mu_8\mu_9}f_{\a\b\c}\,,\
f_{\a\b\c}=f_{(\a\b\c)}
  \end{align*}
  which turns out to be a trivial solution of \eqref{eq5}, i.e. $\Om^{\mathrm{YM}}_{9}=(s+d)\eta_8$ for a total 8-form $\eta_8$. Whether or not \eqref{ym5} and/or \eqref{ym7} are nontrivial solutions of \eqref{eq5} is not completely clear to the author yet. 
  
 (iii) Using the same multi-index notation as above, one can construct further Chern-Simons type solutions 
 of \eqref{eq5} in odd
 dimensions:
\begin{align}
\Dim=2k+1:\ \Om^{\mathrm{CS}}_{2k+1}=
\4\Om_2^{\a_1(\mu_1)}{}_{(\mu_2)}\4\Om_2^{\a_2(\mu_2)}{}_{(\mu_3)}\cdots
\4\Om_2^{\a_k(\mu_k)}{}_{(\mu_{k+1})}\4\Om_1^{\a_{k+1}(\mu_{k+1})}{}_{(\mu_1)}
c_{\a_1\ldots\a_{k+1}}
\label{cs}
  \end{align}
  wherein $c_{\a_1\ldots\a_{k+1}}=c_{[\a_1\ldots\a_{k+1}]}$ if $k=2m$ and 
  $c_{\a_1\ldots\a_{n+1}}=c_{(\a_1\ldots\a_{k+1})}$ if $k=2m+1$. We remark that the 
  Chern-Simons type solution \eqref{cs5} can be written in this form.

\mysection{Consistent deformations in first order formulation}\label{all}

To explore whether or not the consistent first order deformations derived in the previous section exist to all orders we 
employ the first order formulation \cite{Zinoviev:2003ix} of the free theory. The classical fields of that formulation are denoted $\varphi^\a_{\mu\nu\varrho}$ and $\B^\a_{\mu\nu\varrho\sigma}$ whose Lorentz indices have the permutation symmetries
\begin{align}
  \varphi^\a_{\mu\nu\varrho}=-\varphi^\a_{\nu\mu\varrho}\,,\quad
  \B^\a_{\mu\nu\varrho\sigma}=\B^\a_{\mu[\nu\varrho\sigma]}\, .
	\label{f1}
 \end{align}
 We take as Lagrangian of the first order formulation
\begin{align}
  \7{\cal L}^{(0)}=\delta_{\a\b}\,(\tfrac 1{4}\B^\a_{\mu\nu\varrho\sigma}\B^{\b\nu\mu\varrho\sigma}
  -\tfrac 1{4}\B^{\a\mu\nu}\B^\b_{\mu\nu}-\tfrac 1{6}\B^{\a\mu\nu\varrho\sigma}\7\F^\b_{\nu\varrho\sigma\mu}
  +\half \B^{\a\mu\nu}\7\F^\b_{\mu\nu})
	\label{f2}
 \end{align}
 wherein 
 \begin{align}
  \B^\a_{\mu\nu}=\B^{\a\varrho}{}_{\varrho\mu\nu}\,,\quad
  \7\F^\a_{\mu\nu\varrho\sigma}=3\6_{[\mu} \varphi^\a_{\nu\varrho]\sigma}\,,\quad
   \7\F^\a_{\mu\nu}=\7\F^{\a}_{\mu\nu\varrho}{}^\varrho.
	\label{f3}
 \end{align}
The $B$-fields are auxiliary fields which can be eliminated using the algebraic solution of their equations of motion. 
Elimination of the $B$-fields reproduces the Lagrangian \eqref{eq2} (up to a total divergence $\6_\mu R^\mu$) with the definitions\footnote{The fields $\K$ disappear from the Lagrangian upon elimination of the $B$-fields because they contribute only to the total divergence $\6_\mu R^\mu$.}
 \begin{align}
 \varphi^\a_{\mu\nu\varrho}=T^\a_{\mu\nu\varrho}+\K^\a_{\mu\nu\varrho}\,,\quad
 \K^\a_{\mu\nu\varrho}=\varphi^\a_{[\mu\nu\varrho]}\,.
	\label{f4}
 \end{align}
 The ghost fields of the first order formulation of the free theory are denoted $\D^\a_{\mu\nu}$ and 
 $\7H^\a_{\mu\nu\varrho}=\7H^\a_{[\mu\nu\varrho]}$, the ghost-for-ghost fields again $C^\a_\mu$, 
 and the antifields again with a $\star$ and indices corresponding to the indices of the respective field.
 These fields and antifields have the following ghost numbers, antifield numbers, Gra\ss mann parities and BRST transformations (corresponding to the master action $\7\SS^{(0)}=\int [ \7{\cal L}^{(0)}-\sum_\Phi (s\Phi)\Phi^{\star} ] d^\Dim x$):
\begin{equation}
\begin{array}{|c|c|c|c|c|}
 \hline\rule{0em}{2.5ex}
 Z & \mathrm{gh}(Z) & \mathrm{af}(Z) & |Z| &sZ \\
 \hline\rule{0em}{2.5ex}
 \varphi^\a_{\mu\nu\varrho} & 0 & 0 & 0 &2\6_{[\mu}\D^\a_{\nu]\varrho}-\7H^\a_{\mu\nu\varrho}\\
 \hline\rule{0em}{2.5ex}
 \B^\a_{\sigma\mu\nu\varrho} & 0 & 0 & 0 & -\6_{\sigma}\7H^{\a}_{\mu\nu\varrho}\\
 \hline\rule{0em}{2.5ex}
 \D^\a_{\mu\nu}& 1 & 0 & 1 & 6\6_{\mu}\C^\a_{\nu} \\
 \hline\rule{0em}{2.5ex}
  \7\H^\a_{\sigma\mu\nu\varrho} & 1 & 0 & 1 & 0\\
 \hline\rule{0em}{2.5ex}
\C^\a_{\mu}& 2 & 0 & 0 & 0 \\
 \hline\rule{0em}{2.5ex}
 \varphi_\a^{\star\mu\nu\varrho} & -1 & 1 & 1 & 
  \half\delta_{\a\b}\6_\sigma(\B^{\b\varrho\sigma\mu\nu}-3\B^{\b[\sigma\mu}\eta^{\nu]\varrho})=
 \delta_{\a\b}(\7\E^{\b\mu\nu\varrho}-\7\E^{\b[\mu}\eta^{\nu]\varrho})\\
 \hline\rule{0em}{2.5ex}
 \B_\a^{\star\mu\nu\varrho\sigma} & -1 & 1 & 1 & \half \delta_{\a\b}(\B^{\b[\nu\varrho\sigma]\mu}
 -\B^{\b[\nu\varrho}\eta^{\sigma]\mu}
 -\tfrac 13\7\F^{\b\nu\varrho\sigma\mu}
 +\7\F^{\b[\nu\varrho}\eta^{\sigma]\mu})\\
  \hline\rule{0em}{2.5ex}
 \D_\a^{\star\mu\nu}& -2 & 2 & 0 & -2\6_\varrho\varphi_\a^{\star\varrho\mu\nu}\\
    \hline\rule{0em}{2.5ex}
  \7\H_\a^{\star\mu\nu\varrho} & -2 & 2 & 0 &  
  -\varphi_\a^{\star[\mu\nu\varrho]}+\6_\sigma\B_\a^{\star\sigma\mu\nu\varrho}\\
  \hline\rule{0em}{2.5ex}
  \C_\a^{\star\mu} & -3 & 3 & 1 & 6\6_\nu \D_\a^{\star\nu\mu}\\
 \hline
 \end{array}
 \label{f5}
\end{equation}
 wherein
 \begin{align}
  \7\E^\a_{\varrho\sigma\tau\mu\nu}=\6_{[\mu}\B^\a_{\nu]\varrho\sigma\tau}\,,\quad
 \7\E^\a_{\mu\nu\varrho}=-\7\E^\a_{\sigma\mu\nu\varrho}{}^\sigma,\quad
\7\E^\a_{\mu}=\7\E^\a_{\mu\nu}{}^\nu\,.
	\label{f6}
 \end{align}
 We also note that $\D^\a_{\mu\nu}=\S^\a_{\mu\nu}+3\A^\a_{\mu\nu}$, i.e. $\S^\a_{\mu\nu}=\D^\a_{(\mu\nu)}$
 and $\A^\a_{\mu\nu}=\tfrac 13\D^\a_{[\mu\nu]}$.
 
 We now introduce the following total 1-forms and
 2-forms analogously to \eqref{eq13a}:
 \begin{align}
\7\Om_1^{\a\mu\nu\varrho}=\7\H^{\a\mu\nu\varrho}-\B^\a{}_\sigma{}^{\mu\nu\varrho} dx^\sigma,\quad
 \7\Om_2^{\a\mu\nu\varrho}=-\7\E^{\a\mu\nu\varrho}{}_{\sigma\tau} dx^\sigma dx^\tau\label{f7}
 \end{align}
 and the following total $(\Dim-3)$-forms analogously to \eqref{eq13d}:
  \begin{align}
 &\7\Om_{\Dim-3}^{\a\mu}=\sum_{p=\Dim-3}^\Dim\7\om_{\Dim-3-p,p}^{\a\mu}\,,\nonumber\\
  &\7\om_{-3,\Dim}^{\a\mu}=\delta^{ab}\C_\b^{\star\mu} d^\Dim x,\nonumber\\
  &\7\om_{-2,\Dim-1}^{\a\mu}=-6\,\delta^{ab}\D_\b^{\star\nu\mu}(d^{\Dim-1}x)_\nu\,,\nonumber\\
  &\7\om_{-1,\Dim-2}^{\a\mu}=\tfrac{36}{\Dim-3}\delta^{ab}
  (x^{[\nu}\6_{\sigma}\varphi_\b^{\star\varrho\sigma]\mu}
  -x^\tau\eta^{\mu[\nu}\6_{\sigma}\varphi_\b^{\star\varrho\sigma]}{}_\tau
  -\tfrac{2}{\Dim-2}\eta^{\mu[\nu}x^{\varrho}\6_{\sigma}\varphi_\b^{\star\sigma]\tau}{}_\tau)
  (d^{\Dim-2}x)_{\nu\varrho}\,,\nonumber\\
   &\7\om_{0,\Dim-3}^{\a\mu}=-\tfrac{12}{\Dim-4}\, \7\W^{\a\nu\varrho\sigma\mu}{}_\tau x^\tau 
   (d^{\Dim-3}x)_{\nu\varrho\sigma}\label{f8}
 \end{align}
 wherein $\7\W^\a_{\nu\varrho\sigma\mu\tau}$ is defined analogously to $\W^\a_{\nu\varrho\sigma\mu\tau}$
 in \eqref{eq12}, with $\7\E$ in place of $E$. 
The total forms \eqref{f7} and \eqref{f8} fulfill
\begin{align}
(s+d)\7\Om_1^{\a\mu\nu\varrho}= \7\Om_2^{\a\mu\nu\varrho},\quad
(s+d)\7\Om_2^{\a\mu\nu\varrho}= 0,\quad
(s+d)\7\Om_{\Dim-3}^{\a\mu}=0.
\label{f9}
 \end{align}

Therefore solutions $\7\Om^{\mathrm{CM}}_5$, $ \7\Om^{\mathrm{CM}}_7$, 
$\7\Om^{\mathrm{FT}}_5$ and $\7\Om^{\mathrm{CS}}_5$ of equation \eqref{eq5} arise from the solutions
$\Om^{\mathrm{CM}}_5$, $ \Om^{\mathrm{CM}}_7$, 
$\Om^{\mathrm{FT}}_5$ and $\Om^{\mathrm{CS}}_5$
given in equations \eqref{cm5}, \eqref{cm7}, \eqref{ft5} and \eqref{cs5}
by the replacements $\Om_1^{\a\mu\nu\varrho}\rightarrow \7\Om_1^{\a\mu\nu\varrho}$,
$\Om_2^{\a\mu\nu\varrho}\rightarrow \7\Om_2^{\a\mu\nu\varrho}$ and
$\Om_{\Dim-3}^{\a\mu}\rightarrow \7\Om_{\Dim-3}^{\a\mu}$.
We shall show now that the solutions $\7\Om^{\mathrm{CM}}_5$, $ \7\Om^{\mathrm{CM}}_7$, 
$\7\Om^{\mathrm{FT}}_5$ and $\7\Om^{\mathrm{CS}}_5$ are in fact 
equivalent in the cohomology $H(s+d)$ of $(s+d)$ to their respective counterparts
$\Om^{\mathrm{CM}}_5$, $ \Om^{\mathrm{CM}}_7$, 
$\Om^{\mathrm{FT}}_5$ and $\Om^{\mathrm{CS}}_5$, i.e. 
one has
$\Om^{\mathrm{CM}}_5=\7\Om^{\mathrm{CM}}_5+(s+d)\eta_{4}$ for some local total 4-form
$\eta_{4}$ etc.
This follows from the fact that in the first order formulation of the free theory with Lagrangian \eqref{f2} one has
 \begin{align}
\B^\a_{\mu\nu\varrho\sigma}\approx \F^\a_{\nu\varrho\sigma\mu}+\6_\mu\K^\a_{\nu\varrho\sigma}
\label{f10}
 \end{align}
 which implies
  \begin{align}
 \7\E^\a_{\varrho\sigma\tau\mu\nu}\approx  \E^\a_{\varrho\sigma\tau\mu\nu}\,,\quad
 \7\W^\a_{\varrho\sigma\tau\mu\nu}\approx  \W^\a_{\varrho\sigma\tau\mu\nu}\label{f11}
  \end{align}
  and
   \begin{align}
 \7\Om_1^{\a\mu\nu\varrho}+(s+d)\K^{\a\mu\nu\varrho}
 =\H^{\a\mu\nu\varrho}-(\B^\a{}_\sigma{}^{\mu\nu\varrho} -\6_\sigma\K^{\a\mu\nu\varrho})dx^\sigma
 \approx \Om_1^{\a\mu\nu\varrho}
\label{f12}
 \end{align}
 with $\H^{\a\mu\nu\varrho}$ and $\Om_1^{\a\mu\nu\varrho}$ as in \eqref{eq13a}. Hence, in the first 
 order formulation of the free theory the total 1-form
$\7\Om_1^{\prime\,\a\mu\nu\varrho}=\7\Om_1^{\a\mu\nu\varrho}+(s+d)\K^{\a\mu\nu\varrho}$ equals on-shell
the total 1-form $\Om_1^{\a\mu\nu\varrho}$ of \eqref{eq13a}, the 2-form $\7\Om_2^{\a\mu\nu\varrho}$ equals on-shell
the 2-form $\Om_2^{\a\mu\nu\varrho}$ of \eqref{eq13a} and the exterior $(\Dim-3)$-form 
$\7\om_{0,\Dim-3}^{\a\mu}$ of the total $(\Dim-3)$-form
$\7\Om_{\Dim-3}^{\a\mu}$ equals on-shell the exterior $(\Dim-3)$-form 
$\om_{0,\Dim-3}^{\a\mu}$ of the total $(\Dim-3)$-form $\Om_{\Dim-3}^{\a\mu}$ of
\eqref{eq13d}. Therefore the antifield independent parts of the exterior $\Dim$-forms present in 
$\Om^{\mathrm{CM}}_5$, $ \Om^{\mathrm{CM}}_7$, 
$\Om^{\mathrm{FT}}_5$ and $\Om^{\mathrm{CS}}_5$ coincide on-shell
with the respective 
antifield independent parts of the exterior $\Dim$-forms present in the total $\Dim$-forms
$\7\Om^{\prime\,\mathrm{CM}}_5$, $ \7\Om^{\prime\,\mathrm{CM}}_7$, 
$\7\Om^{\prime\,\mathrm{FT}}_5$ and $\7\Om^{\prime\,\mathrm{CS}}_5$
which arise from $\Om^{\mathrm{CM}}_5$, $ \Om^{\mathrm{CM}}_7$, 
$\Om^{\mathrm{FT}}_5$ and $\Om^{\mathrm{CS}}_5$
by the replacements $\Om_1^{\a\mu\nu\varrho}\rightarrow \7\Om_1^{\prime\,\a\mu\nu\varrho}$,
$\Om_2^{\a\mu\nu\varrho}\rightarrow \7\Om_2^{\a\mu\nu\varrho}$ and
$\Om_{\Dim-3}^{\a\mu}\rightarrow \7\Om_{\Dim-3}^{\a\mu}$. As a consequence the solutions of
equation \eqref{eq4} (i.e., the exterior $\Dim$-forms) present in $\Om^{\mathrm{CM}}_5$, $ \Om^{\mathrm{CM}}_7$, 
$\Om^{\mathrm{FT}}_5$ and $\Om^{\mathrm{CS}}_5$ are
equivalent in the cohomology $H(s|d)$ of $s$ modulo $d$ to the respective solutions
present in
$\7\Om^{\prime\,\mathrm{CM}}_5$, $ \7\Om^{\prime\,\mathrm{CM}}_7$, 
$\7\Om^{\prime\,\mathrm{FT}}_5$ and $\7\Om^{\prime\,\mathrm{CS}}_5$ which in turn implies that 
the $(s+d)$-cocycles
 $\Om^{\mathrm{CM}}_5$, $ \Om^{\mathrm{CM}}_7$, 
$\Om^{\mathrm{FT}}_5$ and $\Om^{\mathrm{CS}}_5$ are equivalent in the cohomology $H(s+d)$ 
to the respective $(s+d)$-cocycles
$\7\Om^{\prime\,\mathrm{CM}}_5$, $ \7\Om^{\prime\,\mathrm{CM}}_7$, 
$\7\Om^{\prime\,\mathrm{FT}}_5$ and $\7\Om^{\prime\,\mathrm{CS}}_5$ (i.e. one has
$\Om^{\mathrm{CM}}_5=\7\Om^{\prime\,\mathrm{CM}}_5+(s+d)\eta^\prime_{4}$ for some local total 4-form
$\eta^\prime_{4}$ etc.).\footnote{This follows
by standard arguments from the general feature of the local BRST-cohomology that
the cohomology $H^\Dim_k(\delta|d)$ of the Koszul-Tate differential $\delta$ (= part of $s$ with 
antifield number 1) modulo $d$ vanishes in the space of local exterior $\Dim$-forms which have both positive antifield number $k$
and positive pureghost number, i.e. in the space of exterior $\Dim$-forms which depend at least linearly both on antifields and on fields with positive ghost number, see section 6.3 of \cite{Barnich:2000zw}.} 
Furthermore $\7\Om^{\prime\,\mathrm{CM}}_5$, $ \7\Om^{\prime\,\mathrm{CM}}_7$, 
$\7\Om^{\prime\,\mathrm{FT}}_5$ and $\7\Om^{\prime\,\mathrm{CS}}_5$ are 
equivalent in $H(s+d)$ to
$\7\Om^{\mathrm{CM}}_5$, $ \7\Om^{\mathrm{CM}}_7$, 
$\7\Om^{\mathrm{FT}}_5$ and $\7\Om^{\mathrm{CS}}_5$, respectively, because 
$\7\Om^{\prime\,\mathrm{CM}}_5$, $ \7\Om^{\prime\,\mathrm{CM}}_7$, 
$\7\Om^{\prime\,\mathrm{FT}}_5$ and $\7\Om^{\prime\,\mathrm{CS}}_5$ are all linear
in $\7\Om_1^{\prime\,\a\mu\nu\varrho}$, and because $\7\Om_2^{\a\mu\nu\varrho}$
and $\7\Om_{\Dim-3}^{\a\mu}$ are $(s+d)$-cocycles: e.g., one has
 \begin{align*}
\7\Om^{\prime\,\mathrm{CM}}_5
&=
\epsilon_{\mu_1\ldots\mu_5}\7\Om_{2}^{\a\nu}
\7\Om_2^{\b\mu_1\mu_2\mu_3}
(\7\Om_1^{\c\mu_4\mu_5}{}_\nu+(s+d)\K^{\c\mu_4\mu_5}{}_\nu)\, e_{\a\b\c}\\
&=
\7\Om^{\mathrm{CM}}_5
+(s+d)(\epsilon_{\mu_1\ldots\mu_5}\7\Om_{2}^{\a\nu}
\7\Om_2^{\b\mu_1\mu_2\mu_3}
\K^{\c\mu_4\mu_5}{}_\nu\, e_{\a\b\c}).
 \end{align*}
This implies indeed that $\Om^{\mathrm{CM}}_5$, $ \Om^{\mathrm{CM}}_7$, 
$\Om^{\mathrm{FT}}_5$ and $\Om^{\mathrm{CS}}_5$ are equivalent in $H(s+d)$ to
$\7\Om^{\mathrm{CM}}_5$, $ \7\Om^{\mathrm{CM}}_7$, 
$\7\Om^{\mathrm{FT}}_5$ and $\7\Om^{\mathrm{CS}}_5$, respectively, and that the defomations of the free
theory which arise from these $(s+d)$-cocycles are equivalent as well, respectively.

Now, the first order deformations $\7\SS^{(1)}$ which arise from the solutions $\7\Om^{\mathrm{CM}}_5$, $ \7\Om^{\mathrm{CM}}_7$, 
$\7\Om^{\mathrm{FT}}_5$ and $\7\Om^{\mathrm{CS}}_5$ of \eqref{eq5} 
fulfill $(\7\SS^{(1)},\7\SS^{(1)})=0$ simply because the exterior $\Dim$-forms present in these solutions do not depend on 
the fields $\varphi$, and the only antifields on which these exterior $\Dim$-forms depend are the antifields
$\varphi^\star$ of $\varphi$ (of course, $\Om^{\mathrm{CS}}_5$ and $\7\Om^{\mathrm{CS}}_5$ 
do not depend on antifields at all and therefore it is actually not necessary to substitute $\7\Om^{\mathrm{CS}}_5$ for
$\Om^{\mathrm{CS}}_5$ in order to get $(\SS^{(1)},\SS^{(1)})=0$ for this deformation by itself; however this changes when one considers linear combinations of $\Om^{\mathrm{CM}}_5$, $\Om^{\mathrm{FT}}_5$ and $\Om^{\mathrm{CS}}_5$). Hence, these first order deformations $\7\SS^{(1)}$
provide in fact already a complete deformation $\7\SS=\7\SS^{(0)}+g\7\SS^{(1)}$ of the master action $\7\SS^{(0)}$
of the first order formulation of the free theory. This implies that the first order deformations arising from the
solutions \eqref{cm5}, \eqref{cm7}, \eqref{ft5} and \eqref{cs5} of \eqref{eq5} indeed exist to all orders and the complete deformations in the second order formulation of the free theory with Lagrangian \eqref{eq2}
can be obtained from $\7\SS$ by
eliminating the auxiliary fields $B$ (e.g., perturbatively). It should also be noticed that this reasoning does not only apply
to the Chapline-Manton, Freedman-Townsend and Chern-Simons type solutions in $\Dim=5$
individually but also to any linear combination thereof. 

The author has not found an analogous line of reasoning for 
the Yang-Mills type
deformations yet. The reason is that it does not appear straightforward to find $B$-dependent total forms $\7\Om$ analogous to
\eqref{f7} and \eqref{f8} for the Yang-Mills type deformations which allow a reasoning similar to comment (ii) in section
\ref{first}.

 \mysection{Conclusion}\label{conclusion}
 
 The first order deformations ${\cal L}^{(1)}$ of the Lagrangian \eqref{eq2} that arise from the solutions
 of \eqref{eq5} given in section \ref{first} in dimensions $\Dim=5$ and $\Dim=7$
 are obtained from the antifield independent parts ${\cal L}^{(1)}d^\Dim x$
 of the exterior $\Dim$-forms of these solutions. The first order deformations ${\cal L}^{(1)}_{\mathrm{YM}}$ obtained
 in this way from the solutions \eqref{ym5} and \eqref{ym7} read explicitly
 \begin{align}
\Dim=5:\quad & {\cal L}^{(1)}_{\mathrm{YM}}=\epsilon^{\mu_1\ldots\mu_5}\E^\a_{\mu_1\mu_2\mu_3\sigma\tau}
\F^\b_{\mu_4\nu\varrho}{}^\sigma\F^\c_{\mu_5}{}^{\nu\varrho\tau}f_{\a\b\c}\,, \label{Lym5}\\
\Dim=7:\quad &{\cal L}^{(1)}_{\mathrm{YM}}=\epsilon^{\mu_1\ldots\mu_7}\E^\a_{\mu_1\mu_2\mu_3\sigma\tau}
\F^\b_{\mu_4\mu_5\nu}{}^\sigma\F^\c_{\mu_6\mu_7}{}^{\nu\tau}f_{\a\b\c}\,. \label{Lym7}
  \end{align}
 The first order deformations ${\cal L}^{(1)}_{\mathrm{CM}}$ obtained
 from the solutions \eqref{cm5} and \eqref{cm7} are
 \begin{align}
\Dim=5:\quad & {\cal L}^{(1)}_{\mathrm{CM}}=-12\,\epsilon^{\mu_1\ldots\mu_5}
\W^\a_{\nu_1\ldots\nu_4\varrho} x^\varrho
\E^\b_{\mu_1\mu_2\mu_3}{}^{\nu_1\nu_2}
\F^\c_{\mu_4\mu_5}{}^{\nu_4\nu_3}\, e_{\a\b\c}\,,\label{Lcm5}\\
\Dim=7:\quad &{\cal L}^{(1)}_{\mathrm{CM}}=-4\,\epsilon^{\mu_1\ldots\mu_7}
\W^\a_{\nu_1\nu_2\nu_3\mu_1\varrho} x^\varrho
\E^\b_{\mu_2\mu_3\mu_4}{}^{\nu_1\nu_2}
\F^\c_{\mu_5\mu_6\mu_7}{}^{\nu_3}e_{\a\b\c}\,,\label{Lcm7}
  \end{align}
 and the first order deformations ${\cal L}^{(1)}_{\mathrm{FT}}$ and ${\cal L}^{(1)}_{\mathrm{CS}}$ obtained
 from the solutions \eqref{ft5} and \eqref{cs5} are\footnote{Here we assumed that the
 flat metric has a signature with an odd number of minus signs, such as $(-,+,+,+,+)$. Signatures with an even number of minus signs result in a minus sign in  \eqref{Lcs5} and a plus sign in 
 ${\cal L}_{\mathrm{N}}^{(1)}$ in \eqref{LN}. We remark
 that all results presented in this work are actually valid also for non-Minkowskian metrics, with possible reversed signs in  \eqref{Lcs5} and in ${\cal L}_{\mathrm{N}}^{(1)}$ in \eqref{LN}.}
 \begin{align}
\Dim=5:\quad &{\cal L}^{(1)}_{\mathrm{FT}}=-12\,\epsilon^{\mu_1\ldots\mu_5}\epsilon^{\nu_1\ldots\nu_5}
\W^\a_{\nu_1\nu_2\nu_3\mu_1\varrho} x^\varrho
\W^\b_{\nu_4\nu_5\tau\mu_2\sigma} x^\sigma
\F^\c_{\mu_3\mu_4\mu_5}{}^\tau d_{\a\b\c}\,,\label{Lft5}\\
& {\cal L}^{(1)}_{\mathrm{CS}}=\epsilon^{\mu_1\ldots\mu_5}\epsilon^{\nu_1\ldots\nu_5}
\F^\a_{\mu_1\mu_2\mu_3\nu_1}
\E^\b_{\varrho\sigma\mu_4\nu_2\nu_3}
\E^{\c\varrho\sigma}{}_{\mu_5\nu_4\nu_5}c_{\a\b\c}\,.\label{Lcs5}
  \end{align}
Notice that 
 the first order deformations \eqref{Lym5} and \eqref{Lcm5} exist for any number of Curtright fields (and in particular for only one Curtright field), whereas the first order deformations \eqref{Lcm7} and \eqref{Lft5} require at
 least two Curtright fields, and the first order deformations \eqref{Lym7} and \eqref{Lcs5} require at least three Curtright fields because of equations \eqref{f}, \eqref{e} and \eqref{cd}. Furthermore notice that
 all the above first order deformations are Lorentz invariant, in spite of the explicit $x$-dependence of
the deformations \eqref{Lcm5},  \eqref{Lcm7} and  \eqref{Lft5}.\footnote{This holds because the would-be
infinitesimal Lorentz transformation of $x^\mu$ as a contravariant Lorentz vector vanishes: 
$\xi^\nu\6_\nu x^\mu-x^\nu\6_\nu\xi^\mu=0$ for $\xi^\mu=x^\nu k_\nu{}^\mu$ with constant
$k_\nu{}^\mu$.} This explicit $x$-dependence results from the fact that a gauge invariant improvement
of the conserved exterior $(\Dim-3)$ form  $\lambda_{0,\Dim-3}^{\a\mu}$ in \eqref{eq13} necessarily 
depends explicitly on the coordinates $x$, cf. comment (i) in section \ref{cons}. The deformations 
\eqref{Lcm5},  \eqref{Lcm7} and  \eqref{Lft5} are thus Lorentz invariant but appear to be variant under 
standard spacetime translations. The deformations 
\eqref{Lym5},  \eqref{Lym7} and \eqref{Lcs5} are Poincar\'e invariant.

Notice also that all the above first order deformations are cubic in the Curtright fields and that the
deformations \eqref{Lym5} and \eqref{Lym7} contain four derivatives of the Curtright fields (terms $\6^2T\6\T\6\T$) whereas the 
deformations \eqref{Lcm5}-\eqref{Lcs5} contain five derivatives of the Curtright fields (terms $\6^2T\6^2\T\6\T$), respectively. 

Furthermore, the
deformations \eqref{Lym5}-\eqref{Lft5} of the Lagrangian are accompanied by deformations of the gauge transformations
of the free theory. The first order deformations of the gauge transformations are obtained from the corresponding solutions of \eqref{eq5} given in section \ref{first}, more precisely from the terms with antifield number 1 in the
exterior $\Dim$-forms of these solutions. We leave it to the interested reader to write out these
deformations of the gauge transformations explicitly. The commutator algebra of the first order deformed gauge transformations
remains Abelian in all cases, however. This corresponds to the fact that the exterior $\Dim$-forms of the solutions
 of \eqref{eq5} given in section \ref{first} do not contain terms with antifield number exceeding 1.
 
The deformations derived here are thus compatible with the results of \cite{Bekaert:2002uh,Bekaert:2004dz} where it was shown that Poincar\'e invariant first order consistent deformations of the free theory that modify nontrivially the gauge transformations leave the
commutator algebra of the deformed gauge transformations Abelian on-shell, and that there are actually no nontrivial
consistent deformations of this type containing at most three derivatives of the Curtright fields.
In fact it can easily be shown that $x$-independent and Lorentz invariant nontrivial consistent deformations that are strictly invariant under the
gauge transformations
of the free theory and contain at most four derivatives do not exist either. Indeed, according to the results of
\cite{Bekaert:2002uh,Bekaert:2004dz} such deformations can be taken to 
be quadratic in the tensors $\E^{\a\mu\nu\varrho\sigma\tau}$ but all such quadratic terms actually vanish on-shell up to a
total divergence because of \eqref{eq10}-\eqref{eq12b} and are thus trivial deformations of the Lagrangian \eqref{eq2}. Therefore it seems that the above deformations might
actually provide the simplest possible Lorentz invariant nontrivial deformations of the free theory 
in dimensions $\Dim=5$ and $\Dim=7$ at first order.

As shown in section \ref{all} the above first order deformations \eqref{Lcm5}-\eqref{Lcs5} can in fact be extended to all orders, most
readily using the first order formulation of the theory. Furthermore in $\Dim=5$ any linear
combination of the deformations \eqref{Lcm5}, \eqref{Lft5} and \eqref{Lcs5} can be extended to all orders.
Whether or not the first order deformations \eqref{Lym5} and \eqref{Lym7} can be extended to higher orders is left open here.

We also remark that in all above first order deformations the tensors $\E^{\a\mu\nu\varrho\sigma\tau}$ can be
replaced by the traceless tensors $\W^{\a\mu\nu\varrho\sigma\tau}$ \eqref{eq12} and vice versa because of
$\E^{\a\mu\nu\varrho\sigma\tau}\approx\W^{\a\mu\nu\varrho\sigma\tau}$, see also remark (iii) in section \ref{cons} (such replacements provide equivalent deformations and modify the deformed gauge transformations).

The author admits that he has no complete proof yet that the above deformations are really nontrivial. Therefore
some (or all) of these deformations may actually turn out to be trivial. 
The proof of nontriviality is hampered by the possible explicit $x$-dependence of
the terms (forms) that may make the deformations trivial. The author plans to investigate this issue, and whether or not
the first order deformations \eqref{Lym5} and \eqref{Lym7} can be extended to higher orders in a future work
(unless someone else does the job).\footnote{In particular there seems to be no obvious counterpart of the Yang-Mills type self-interactions  \eqref{Lym5} and \eqref{Lym7} in previous works, such as \cite{Metsaev:2005ar} which aimed to classify cubic consistent interactions of ``mixed symmetry'' and higher spin fields rather completely. Therefore it seems to be worthwhile to check whether or not especially the self-interactions \eqref{Lym5} and \eqref{Lym7} are trivial, and if they turn out to be nontrivial, to clarify their relation to results of previous works.}
However, the similarity of 
\eqref{Lym5}-\eqref{Lcs5} to
Yang-Mills \cite{Yang:1954ek}, Chapline-Manton \cite{Chapline:1982ww}, Freedman-Townsend \cite{Freedman:1980us}
and Chern-Simons \cite{Chern:1974ft} interactions, respectively, in combination with
some BRST-cohomological considerations, suggests the nontriviality of the deformations. 

Let me therefore briefly comment on similarities (and differences) of the deformations 
\eqref{Lym5}-\eqref{Lcs5}  to
Yang-Mills, Chapline-Manton, Freedman-Townsend and Chern-Simons interactions. To that end standard $p$-form gauge potentials
are denoted $A^\a_p=\tfrac 1{p!}A^\a_{\mu_1\ldots\mu_p}dx^{\mu_1}\ldots dx^{\mu_p}$, the corresponding field strength $(p+1)$-forms
$\F^\a_{p+1}=dA^\a_p$ and the Hodge duals of the field strength forms $\5\F^\a_{\Dim-p-1}$ . 

Yang-Mills interactions
in $\Dim$ dimensions are $\5\F^\a_{\Dim-2}A^\b_1A^\c_1f_{\a\b\c}$. This is analogous
to \eqref{ym5} and \eqref{ym7} with $\Om_{\Dim-2}^{\a\cdots}$ 
corresponding to $\5\F^\a_{\Dim-2}$,
and $\Om_1^{\b\cdots}$ and $\Om_1^{\c\cdots}$ 
corresponding 
to $A^\b_1$ and $A^\c_1$, respectively. I stress that 
the terminology ``Yang-Mills type interactions'' used in the present work only relates to this structure of the interactions and not to the commutator algebra of the deformed gauge transformations
(i.e. it is not related to the question whether or not this algebra is Abelian). 

Cubic Chapline-Manton interactions in $\Dim$ dimensions with two 1-form gauge fields are
$\5\F^\a_{\Dim-3}\F^\b_2 A^\c_1 e_{\a\b\c}$. This is analogous
to \eqref{cm5} and \eqref{cm7} with 
$\Om_{\Dim-3}^{\a\cdot}$ 
corresponding to $\5\F^\a_{\Dim-3}$,
$\Om_2^{\b\cdots}$ 
corresponding 
to $F^\b_2$, and 
$\Om_1^{\c\cdots}$ 
corresponding 
to $A^\c_1$.

Cubic Freedman-Townsend interactions in $5$ dimensions are
$\5\F^\a_{1}\5\F^\b_{1}\A^\c_{3} d_{\a\b\c}$. This is analogous
to \eqref{ft5} with  $\Om_2^{\a\cdot}$ and $\Om_2^{\b\cdot}$ 
corresponding to $\5\F^\a_{1}$
and $\5\F^\b_{1}$, and
$\Om_1^{\c\cdots}$ 
corresponding to $\A^\c_{3}$. 
The correspondence here does not match the form-degrees and total degrees
but concerns the structure $\5\F\5\F\A$.

Cubic Chern-Simons interactions in $5$ dimensions are
$\A^\a_{1}\F^\b_{2}\F^\c_{2} c_{\a\b\c}$. This is analogous
to \eqref{cs5} with $\Om_1^{\a\cdots}$ 
corresponding 
to $A^\a_1$, and $\Om_2^{\b\cdots}$ and $\Om_2^{\c\cdots}$ 
corresponding 
to $F^\b_2$ and $F^\c_2$.

The difference of the deformations 
\eqref{Lym5}-\eqref{Lcs5}
as compared to standard Yang-Mills, Chapline-Manton, Freedman-Townsend and Chern-Simons interactions
results on the one hand from the additional Lorentz indices of the $\Om$'s as compared
to standard $p$-form gauge potentials $A_p$ and, on the other hand, from the fact that the action 
$\int {\cal L}^{(0)}d^\Dim x$ does not correspond to the
standard Maxwell type action for free $p$-form gauge potentials $A_p$ containing terms $\int\F_{p+1} \5\F_{\Dim-p-1}$.

As far as the author knows the self-interactions of Curtright fields obtained in this paper 
have not been disclosed anywhere else in the literature so far. 
Nevertheless, self-interactions of ``mixed symmetry gauge fields'' similar to the Chapline-Manton type interactions 
\eqref{Lcm5} and \eqref{Lcm7} have been found in \cite{Bekaert:2004dz}. They are disclosed under item (iv) in
section 8.1 of the arXiv-version of  \cite{Bekaert:2004dz}. The self-interactions disclosed there also depend explicitly on the coordinates $x$ and have a structure analogous to the Chapline-Manton type interactions 
\eqref{Lcm5} and \eqref{Lcm7}. In the particular case $(p,q)=(2,1)$ (corresponding to a Curtright field) and $s=1$ (using the notation of \cite{Bekaert:2004dz}) the 
interactions given there will very likely in $\Dim=5$ provide a self-interaction of a Curtright field equivalent to  the
Chapline-Manton type interaction 
\eqref{Lcm5} (for one Curtright field) when the Lorentz structure of the fields is taken into account.\footnote{Section 8.1 of \cite{Bekaert:2004dz} actually concerns the cases $k>1$ in the notation used there, i.e. deformations which may lead to deformed gauge transformations with a non-Abelian commutator algebra. An interaction with
 $(p,q)=(2,1)$ and $s=1$ however actually represents the case $k=1$, i.e. it corresponds to a deformation which leaves the commutator algebra of the deformed gauge transformations Abelian at first order. This is compatible with the results of the present paper.}

 Let me finally remark that it is quite straightforward to construct interactions of 
 Curtright fields with other fields in appropriate dimensions similar to the above self-interactions using the approach of the present paper. For instance, similarly to
 equation \eqref{ft5} one easily constructs solutions $\Om^{\mathrm{N}}_5$ of equation \eqref{eq5} in $\Dim=5$ which
provide first order deformations ${\cal L}_{\mathrm{N}}^{(1)}$ of the Lagrangian from the total $(\Dim-3)$-forms \eqref{eq13d} for
 $\Dim=5$ and 
 the total 1-form $\Om_1=C+A_\mu dx^\mu$ which is the sum of a standard Abelian 1-form gauge potential 
 $A_\mu dx^\mu$ and the corresponding ghost field $C$:
\begin{align}
 \Dim=5:\ &\Om^{\mathrm{N}}_5=
\Om_{2}^{\a\mu}\Om_{2\mu}^{\b} \Om_1 g_{\a\b}\,,\
\Om_1=C+A_\mu dx^\mu,\
  {\cal L}_{\mathrm{N}}^{(1)}=-A_\mu j^\mu,\nonumber\\
& j^\mu=\epsilon^{\mu\nu_1\nu_2\varrho_1\varrho_2}\4W^\a_{\nu_1\nu_2\sigma}
\4W^\b_{\varrho_1\varrho_2}{}^\sigma g_{\a\b},\
\4W^\a_{\nu_1\nu_2\sigma}=\epsilon_{\nu_1\ldots\nu_5}\W^{\a\nu_3\nu_4\nu_5}{}_{\sigma\varrho}x^\varrho
\label{LN}
\end{align}
wherein $ g_{\a\b}=g_{\b\a}$ are constant symmetric coefficients and
${\cal L}_{\mathrm{N}}^{(1)}$ is a Noether coupling of the gauge field $A_\mu$ and
an (``improved'') Noether current $j^\mu$ of the free theory
($\6_\mu j^\mu\approx 0$). Analogously one constructs in $\Dim=5$ Chern-Simons type interactions of Curtright fields and
a standard Abelian 1-form gauge potential from the solution 
$\Om_{2}^{\a\mu\nu\varrho}\Om_{2\mu\nu\varrho}^{\b} \Om_1 k_{\a\b}$ of \eqref{eq5} wherein 
$k_{\a\b}=k_{\b\a}$ are constant symmetric coefficients and $ \Om_{2}^{\a\cdots}$ and  $ \Om_{2}^{\b\cdots}$ are
the 2-forms of \eqref{eq13a}.
Cubic interactions $\6\T\6\T\6^2 h$ of a Curtright field $\T$ with a symmetric 2-tensor field $h_{\mu\nu}=h_{\nu\mu}$ representing the
 metric field of linearized general relativity  were obtained in section 5 
 of \cite{Boulanger:2011qt} (see equation (5.14) there). These interactions are reminiscent of the Yang-Mills type self-interactions \eqref{Lym5} and \eqref{Lym7} and may be constructible analogously to \eqref{ym5} and\eqref{ym7}
 using a total curvature $(\Dim-2)$-form for the $h$-field in place of $\Om_{\Dim-2}^{\a\mu_1\mu_2\mu_3}$. This indicates that the approach used here may also be useful for the construction of consistent
 interactions of other ``mixed symmetry'' or higher spin fields. Cubic interactions of various fields of that type in various dimensions, both in flat space and in anti-de Sitter space, were constructed by different methods in \cite{Metsaev:1993gx,Metsaev:1993mj,Metsaev:2005ar,Boulanger:2011se,Boulanger:2012dx}, amongst others (see also references cited therein).
 
 {\bf Acknowledgement:}
The author thanks Nicolas Boulanger for correspondence. 

\end{document}